\documentclass[adp,a4paper,fleqn%
 ,finallayout
]{w-art}
\usepackage{times,cite,w-thm}
\theoremstyle{plain}

\theoremstyle{definition}

\usepackage[]{graphicx}
\usepackage{mathbbol}
\providecommand{\sqs}{\Delta(\omega)}
\providecommand{\ca}[1]{c^{\ast}_{#1}}
\providecommand{\fa}[1]{f^{\ast}_{#1}}
\providecommand{\tris}{\Delta^{\text{N}}(\omega)}
\providecommand{\sks}[1]{\sum_{#1}}
\providecommand{\sis}[1]{\sum_{#1}}
\def\bvec#1{\boldsymbol{#1}}
\providecommand{\ket}[1]{\ensuremath{|#1\rangle}}
\providecommand{\bra}[1]{\ensuremath{\langle #1 |}}
\newcommand\Let{\mathrel{\mathop:\!\!=}}

\providecommand{\iv}{\bvec{i}}
\providecommand{\one}[1]{\ensuremath{| #1\rangle\langle #1| }}
\providecommand{\lr}[1]{\ensuremath{\langle #1 \rangle}}
\providecommand{\Rfrac}[8]{\ensuremath{\frac{\mathcal{R}_{#1 #2}(E_{#3},E_{#4})}{E_{#5 #6}(i\omega-E_{#7 #8})}}}
\providecommand{\Qfrac}[8]{\ensuremath{\frac{\mathcal{Q}_{#1 #2}(i\omega,E_{#3},E_{#4})}{(i\omega-E_{#5 #6})(i\omega-E_{#7 #8})}}}
\providecommand{\Qfracm}[8]{\ensuremath{\frac{\mathcal{Q}_{#1 #2}(-i\omega,E_{#3},E_{#4})}{(i\omega-E_{#5 #6})(i\omega-E_{#7 #8})}}}

\begin{document}
\pagespan{1}{}



\title[Superperturbation theory on the real-axis]{Superperturbation theory on the real-axis}


\author[C. Jung]{Christoph Jung\inst{1}%
  \footnote{E-mail:~\textsf{cjung@physnet.uni-hamburg.de}}}
\address[\inst{1}]{1. Institut f\"ur theoretische Physik\\Universit\"at Hamburg\\Jungiusstra\ss e 9\\22355 Hamburg Germany}
\address[\inst{2}]{Centre de Physique Th\'eorique (CPHT)\\ \'Ecole Polytechnique\\91128 Palaiseau Cedex, France}
\author[A. Wilhelm]{Aljoscha Wilhelm\inst{1}}
\author[H. Hafermann]{Hartmut Hafermann\inst{2}}
\author[S. Brener]{Sergey Brener\inst{1}}
\author[A. Lichtenstein]{Alexander Lichtenstein\inst{1}}
\begin{abstract}
In this article we formulate the superperturbation theory for the Anderson impurity model on the real axis. The resulting impurity solver allows to evaluate dynamical quantities without numerical analytical continuation by the maximum entropy method or Pad\'e approximants. This makes the solver well suited to study multiplet effects in solids within the dynamical mean field theory. First examples including multi-orbital problems are discussed.
\end{abstract}

\maketitle

\section{Introduction}
The investigation of strongly correlated materials is one of the main challenges of modern condensed matter theory, which has inspired the research activities of many scientists in the last decades. The physical properties of such materials are characterized by an interplay between the Coulomb interaction of the nearly confined electrons and their kinetic energy.
This competition results in a delicate balance of localization and delocalization, which requires accurate non-perturbative approaches. Dynamical mean-field theory (DMFT) has become a standard tool for the investigation of strongly correlated materials \cite{kotliar:53,Georges:1996tv} and has been applied to models as well as to the investigation of real materials\cite{Lichtenstein:1998kx,Held:2007vn,Kotliar:2006fk}. DMFT for the first time allowed to treat the coherent low-energy excitations and the high-energy excitations as well as their mutual feedback on the same footing.\\
The main concept of the DMFT is to replace the correlated lattice by a single impurity embedded in a self-consistent effective medium. In contrast to classical mean-field approaches, the effective medium in DMFT is represented by an energy dependent electronic bath which takes local temporal quantum fluctuations into account, whereas spatial fluctuations are frozen out.
The solution of the DMFT equations in turn requires accurate impurity solvers. Quantum Monte Carlo (QMC) algorithms are nowadays widely used for this purpose. 
Continuous-time QMC solvers\cite{Rubtsov:2005lq,Werner:2006px} allow to tackle problems with $5$ or even $7$ orbitals as required for systems with open $d$ or $f$-shells, respectively. Nevertheless, they suffer from two major drawbacks: The fermionic sign problem for a general Coulomb vertex and --more severely-- the fact that QMC algorithms work in the imaginary time domain require the numerically ill-conditioned analytical continuation of stochastic data to the real axis. This makes it difficult to reliably access spectral properties and to study e.g. multiplett effects in solids.\\
In principal a rigorous approach, which can work on the real-axis, is the exact diagonalization (ED) or Lanczos  scheme\cite{CAFFAREL:1994ly}. Here the  continuous bath is discretized by a small collection of bath sites, so that the problem can be diagonalized exactly.
Due to the exponential growth of the Hilbert space with the number of bath parameters, one is limited to as little as one or two bath sites per \emph{d}-orbital, so that finite size effects may become dominant. In addition, an ambiguity arises in determining the effective bath parameters to find an optimal representation of the input hybridization.
Alternative approximate approaches comprise the iterated perturbation theory (IPT) \cite{PTPS.46.244,PTP.53.1286}, the fluctuation-exchange (FLEX) \cite{BICKERS:1989uq} or non-crossing approximation (NCA) \cite{Keiter:1971fk} and the related hybridization expansion \cite{Dai:2005qd}. These perturbative methods are naturally limited to some parameter window and fail outside this region. FLEX and IPT are applicable for weak-coupling, while the hybridization expansion works in the strong coupling regime.\\
Recently an impurity solver has been developed by combining ED with a diagrammatic approach. The key idea is to formulate a perturbation expansion around the ED solution by employing a transformation to auxiliary, so-called dual fermions\cite{Rubtsov:2008zr}. We refer to the perturbation of a non-trivial (i.e. interacting) albeit numerically solvable reference problem as a superperturbation\cite{Hafermann:2009sf}.
It has been shown that the method becomes exact in two opposite limits: For weak coupling and and strong hybridization, the approach becomes equivalent to a standard perturbation expansion in the interaction. In the opposite limit of strong interaction and weak hybridization the formalism resembles the strong coupling expansion\cite{Dai:2005qd}.
This method can be sytematically improved by either including more bath sites in the ED (which essentially decreases the perturbation) or by including more diagrams.

In the previous work the solver has been formulated in imaginary time. Analytical continuation was found to be significantly more stable than in QMC due to the absence of statistical noise. 
In this article, we formulate the superperturbation theory on the real-axis, making analytical continuation from intermediate imaginary time results obsolete.

\section{Superperturbation formalism}

To set the stage, we briefly review the superperturbation formalism in the following. For further reading we refer the interested reader to\cite{Hafermann:2009sf}.
The model under consideration is the Anderson impurity model described by the following Hamiltonian:
\begin{equation}\label{eq:Ham_init}
H=\sum_{k\gamma}\epsilon^{b}_{k\gamma}b^{\dagger}_{k\gamma}b_{k\gamma} +\sum_{\alpha}\epsilon^{c}_{\alpha}c^{\dagger}_{\alpha}c_{\alpha} +H_{\text{int}}[c^{\dagger},c]
+\sum_{k\alpha\beta}\left( V_{k}^{\alpha\beta}c_{\alpha}^{\dagger}b_{k\beta}+V^{\ast\beta\alpha}_{k}b_{k\alpha}^{\dagger}c_{\beta}\right).
\end{equation}
Here Greek letters are used as a combined index for orbital and spin degrees of freedom. $b^{\dagger}$ ($c^{\dagger}$) and $b$ ($c$) are the bath (impurity) creation and annihilation operators, respectively. $H_{\text{int}}$ is the local electron-electron interaction, $V_{k}^{\alpha\beta}$ are the transition amplitudes for hopping processes from the bath to an impurity orbital. To derive a dual formulation of the problem, we first integrate out the bath degrees of freedom, which leads to the conventional action representation:
\begin{equation}\label{eq:deriv_init_action}
S[\ca{},c]=-\sum_{\omega \alpha \beta} \ca{\omega \alpha}[(i\omega+\mu)\Eins - \sqs]_{\alpha \beta}c_{\omega \beta}+ S^{\text{int}}[\ca{},c_{}].
\end{equation}
Here $S^{\text{int}}$ is a non-gaussian interaction term and 
$$[\sqs]_{\alpha\beta} = \sum_{k\gamma} \frac{V^{\alpha\gamma}_{k} (V^{\beta\gamma}_{k})^*}{i\omega - \epsilon_{k\gamma}}  $$ 
is the hybridization function of the full system. The first step in the formulation of the superperturbation is to express the action of the model in terms of that of an exactly solvable reference problem and a difference term. To this end, we add and subtract a hybridization function $\tris$ corresponding to a discrete bath:
\begin{align}\label{deriv_new_ac}
S[\ca{},c]&=S^{\text{ref}}[\ca{},c] -\sks{\omega\alpha \beta}\ca{\omega \alpha}[\tris-\sqs]_{\alpha \beta}c_{\omega \beta},\\\label{deriv_ref_ac}
S^{\text{ref}}[\ca{},c]&=-\sis{\omega\alpha \beta}\ca{\omega \alpha}[(i\omega+\mu)\Eins-\tris]_{\alpha \beta} c_{\omega \beta}+ S^{\text{int}}[\ca{},c_{}].
\end{align}
The reference system with a discrete bath with $\text{N}$ bath sites shares the interaction part of the original problem and can be solved efficiently by exact diagonalization. Note that this step leaves the hybridization $\tris$ unspecified. 

\begin{figure}[t]
\sidecaption
 \includegraphics[scale=0.6]{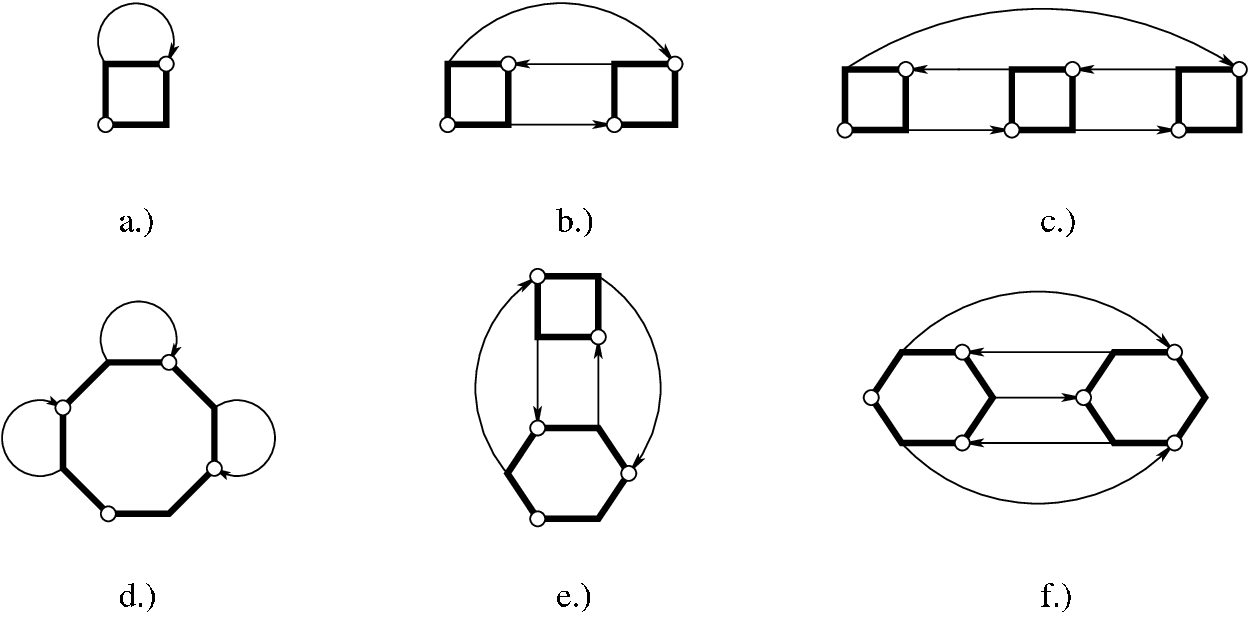}
 \caption{First few diagrams contributing to the dual self-energy. Boxes are the reducible vertices of the reference system ($\gamma^{(4)}$ and $\gamma^{(6)}$), lines denote the dual propagator ($G^{\text{d}}_{0}$).}\label{fig:dual_diag}
\end{figure}

The second step is to reformulate the problem in such a way that a perturbative treatment of the difference term $D(\omega)=\tris-\sqs$ can be performed.
Since $S^{\text{ref}}$ contains a non-quadratic part Wick's theorem is not directly applicable. Therefore, we introduce auxiliary (dual) fermionic degrees of freedom using an exact Gaussian identity in the path integral:
\begin{equation}
e^{\ca{\alpha}a_{\alpha \beta}b^{-1}_{\beta \gamma}a_{\gamma \delta}c_{\delta}}=\frac{1}{\det{b}}\int \mathcal{D}[\fa{},f]e^{-\fa{\alpha}b_{\alpha \beta}f_{\beta}+\fa{\alpha}a_{\alpha \beta}c_{\beta}+\ca{\alpha}a_{\alpha \beta}f_{\beta}},
\end{equation}
where the matrices $a$ and $b$ have the following form:
\begin{equation} \left.\begin{aligned}
a&=-g^{-1}(\omega)\\
b&=g^{-1}(\omega)[\tris-\sqs]^{-1}g^{-1}(\omega)\end{aligned} \right\} \rightarrow a_{\alpha \beta}b^{-1}_{\beta \gamma}a_{\gamma \delta}=[\tris-\sqs]_{\alpha \delta}=D(\omega), 
\end{equation}
with $g(\omega)$ being the exact single particle Green's function of the reference system Eq. \eqref{deriv_ref_ac}. After the transformation, the resulting action has a mixed representation of $c$- and $f$-fermions:
\begin{align}
S[\ca{},c,\fa{},f]=&S^{\text{ref}}[\ca{},c_{}]+S^{\text{c}}[\fa{},f,\ca{},c_{}]+\sks{\omega\alpha \beta}\fa{\omega \alpha}[g(\omega)D(\omega)g(\omega)]^{-1}_{\alpha \beta}f_{\omega \beta},
\end{align}
where the coupling between the $c$ and $f$-fermions is given by
\begin{align}
S^{\text{c}}[\fa{},f,\ca{},c]=&\sis{\omega \alpha \beta}\fa{\omega \alpha}g^{-1}_{\alpha \beta}(\omega)c_{\omega \beta} + \sis{\omega \alpha \beta}\ca{\omega \alpha}g^{-1}_{\alpha \beta}(\omega)f_{\omega \beta}.\label{eq:SC}
\end{align}
The original fermionic degrees of freedom can formally be integrated out exactly. To this end, we expand in $\exp (-S^{c})$ in the path integral representation of the partition function. It is convenient to reexpress the result in the following form:
\begin{equation}\label{eq:derivation:define_dual_pot}
\begin{split}
\int \exp(-S^{\text{ref}}[\ca{},c]-S^{\text{c}}&[\ca{},c,\fa{},f])\mathcal{D}[\ca{},c]\\
 &\overset{!}{=}Z_{\text{ref}}\exp(-\sis{\omega\alpha \beta}\fa{\omega \alpha}g^{-1}_{\alpha \beta}(\omega)f_{\omega \beta} + V[\fa{},f]).
 \end{split}
\end{equation}
This equation defines the dual potential, which gathers two-particle and higher-order interaction terms. The key point is that due to the presence of $S^{\text{ref}}$ on the left-hand-side of (\ref{eq:derivation:define_dual_pot}), integrating out the original fermions corresponds to performing the average over the degrees of freedom of the reference system: $Z_{\text{ref}}\langle\ldots\rangle_{\text{ref}}\Let \int\ldots\mathcal{D}[\ca{},c]$. The resulting dual action has the following form:
\begin{equation}\label{eq:derivation:dual_action_with_exp}
 S^{\text{d}}[\fa{},f]=-\sks{\omega\alpha \beta}\fa{\omega \alpha}[G^{\text{d}}_{0}(\omega)]^{-1}_{\alpha \beta}f_{\omega \beta}+V[\fa{},f].
\end{equation}
As a result of averaging, both the bare dual matrix Green function $G^{\text{d}}_{0}(\omega)=-g(\omega)[g(\omega)+D(\omega)^{-1}]^{-1}g(\omega)$ and the dual potential
\begin{equation}
V[\fa{\iv},f_{\iv}]=-\frac{1}{4}\gamma^{(4)}_{\alpha \beta \gamma \delta}\fa{\alpha}f_{\beta}\fa{\gamma}f_{\delta}+\frac{1}{36}\gamma^{(6)}_{\alpha \beta \gamma \delta \mu \nu}\fa{\alpha}f_{\beta}\fa{\gamma}f_{\delta}\fa{\mu}f_{\nu} \mp \dots \,.
\end{equation}
contain the correlation functions of the reference system: $g(\omega)$ denotes its single-particle Green's function and $\gamma^{(n)}$ are the corresponding reducible vertices. The two-particle vertex for example is given by:
\begin{equation}\label{eqn:gamma4}
 \gamma^{(4)}_{\alpha \beta \gamma \delta}=g^{-1}_{\alpha \alpha'}g^{-1}_{\gamma \gamma'}[\chi^{\text{Ref}}_{\alpha' \beta' \gamma' \delta'}-\chi^{0,\text{Ref}}_{\alpha' \beta' \gamma' \delta'}]g^{-1}_{\beta' \beta}g^{-1}_{\delta' \delta},
\end{equation}
with $\chi^{\text{ref}}$ being the two-particle Green's function of the reference system and $\chi^{0, \text{ref}}$ its unconnected part.\\
These quantities can be calculated straightforwardly on Matsubara frequencies from the Lehmann representation of the single- and two-particle Green's functions \cite{Hafermann:2009sf}.
So far, \eqref{eq:derivation:dual_action_with_exp} is only a reformulation of the initial action in Eq. \eqref{eq:deriv_init_action}. 
The dual problem can now be treated perturbatively, which essentially corresponds to an expansion around the reference problem. 
For example the self-energy correction stemming from the first diagram in Fig. \ref{fig:dual_diag} is given by
\begin{equation}\label{eqn:sigmaa}
(\Sigma^{d}_{(a)})_{\alpha\beta} = -\gamma^{(4)}_{\alpha\beta\gamma\delta} (G^{d}_{0})_{\delta\gamma}.
\end{equation}
After summing up a certain class of diagrams, we obtain a physical result (i.e. solution in terms of the physical c-fermion Green function $G$) by transforming the dual Green's function back to c-fermions using the following exact relation \cite{Rubtsov:2008zr}:
\begin{align}\label{eq:back_trafo}
G(\omega)=D(\omega)^{-1}+[g(\omega)D(\omega)]^{-1} G^{\text{d}}(\omega)[D(\omega)g(\omega)]^{-1}.
\end{align} 

It can be shown that the dual perturbation theory becomes equivalent to conventional perturbation theory in the limit of small interaction and strong hybridization. It is instructive to consider its behavior in the opposite strong coupling limit. For an expansion around the atomic limit, i.e. $\tris\equiv 0$, combining equations \eqref{eqn:gamma4}, \eqref{eqn:sigmaa} and \eqref{eq:back_trafo} with the lowest order approximation to the dual Green's function, $G^{\text{d}}\approx  G^{\text{d}}_{0} + G^{\text{d}}_{0}\Sigma^{\text{d}}_{(a)} G^{\text{d}}_{0}$ leads to the following expression:
\begin{equation}\label{eqn:ghybexpansion}
G_{\alpha\beta} \approx g_{\alpha\beta} + g_{\alpha\beta} \beta Tr[g\Delta] - \chi_{\alpha\beta\gamma\delta}\Delta_{\delta\gamma},
\end{equation}
which recovers the result obtained from an expansion of the imaginary time Green function up to first order in the hybridization, as given in Ref.\cite{Dai:2005qd}.

The reference system, which is specified through the hybridization function $\tris$, should be chosen in an optimal way. Here this is even more crucial than in conventional ED, since for the present approach the number of bath sites should be kept at a minimum to increase the efficiency. The choice of $\tris$ directly affects the perturbation $D(\omega)$, which in a certain sense should be minimal. This can be achieved by minimizing a predefined distance function\cite{Koch:2008vp,Georges:1996tv}.\\

\section{Reformulation on the real axis}\label{chap5:analytical_con}

In order to access spectral properties, the previous formulation required analytical continuation of the imaginary time data. This is an ill-posed problem, and different methods have been developed for this purpose. 
The maximum entropy method (MAXENT) \cite{PhysRevB.44.6011} was specifically designed for inferring spectral properties from statistical data using Bayesian methods. Very fine structures, like multiplets for example, are very hard to resolve using this method. In the Pad\'e approach\cite{Vidberg:1977hc} the function of interest is approximated by a rational function. Pad\'e can give accurate information on the real axis, provided it is applied to noise-free input data. The approximation through a rational function however is unneccessary for the superperturbation approach and may introduce spurious features. It order to extract unbiased real axis information, it is desirable to formulate this approach directly on the real axis.

In ED, real axis information is readily obtained by performing the substitution $i\omega\to E+i\delta$ in the Lehmann representation, where $\delta$ is a small broadening parameter. For the single-particle Green's function this leads to:
\begin{equation}
 g_{\alpha\beta}(E+i\delta)=\frac{1}{Z}\sum_{n,m}\frac{\bra{n}c_{\alpha}\one{m}c^{\dagger}_{\beta}\ket{n}}{E+i\delta + E_n -E_m}\left( e^{-\beta E_n}+e^{-\beta E_m}\right).
\end{equation}
From latter expression the density of states (DOS) is obtained as $DOS(E)=-1/\pi \text{Im}\, g(E+i\delta)$. 
In the superperturbation, analytical continuation is more involved, as one has to take care of the individual diagrams. In the following we discuss the analytical continuation of the first diagram (diagram a.) of Fig. \ref{fig:dual_diag}). This diagram has been found to yield by far largest correction to the initial solution.
Inserting \eqref{eqn:gamma4} into \eqref{eqn:sigmaa}, we obtain
\begin{equation}
\Sigma^{\text{d}(a)}_{\alpha \beta}=-g_{\alpha \beta}^{-1}\beta\text{Tr} [\Delta^{\text{d}}g]+\Delta^{\text{d}}_{\alpha \beta}-g^{-1}_{\alpha \alpha'} \chi_{\alpha' \beta' \gamma' \delta'} g^{-1}_{\beta' \beta}\Delta^{\text{d}}_{\delta' \gamma'},\label{eq:sigma_dai}
\end{equation}
which is very similar to expression \eqref{eqn:ghybexpansion}, where now $\Delta^{\text{d}}\Let [g+(\Delta^{\text{N}}-\Delta)^{-1}]^{-1}$ plays the role of $\Delta$.

While $g$ and $\Delta^{\text{N}}$ are readily accessed on the real axis , we can make use of a result of Ref.\cite{Dai:2005qd} which allows to calculate the product of the two-particle Green function with $\Delta^{d}$ directly within ED.
\begin{equation}
\begin{split}
\frac{1}{\beta}\sum_{i\omega_{n'}}&\chi_{\alpha \beta \gamma \delta}(i\omega,i\omega,i\omega',i\omega')\Delta^{\text{d}}_{\delta \gamma}(i\omega')=\sum_{\gamma \delta}\sum_{ijkl}\times\\
+&\lr{c_{\alpha}\ca{\gamma}c_{\delta}\ca{\beta}}_{ijkl}\bigl[\Rfrac{\gamma}{\delta}{j}{k}{j}{l}{j}{i}+\Rfrac{\gamma}{\delta}{l}{k}{l}{j}{l}{i}+\Qfrac{\gamma}{\delta}{i}{k}{l}{i}{j}{i}\bigr]\\
+&\lr{c_{\alpha}c_{\delta}\ca{\gamma}\ca{\beta}}_{ijkl}\bigl[\Rfrac{\gamma}{\delta}{k}{j}{j}{l}{j}{i}+\Rfrac{\gamma}{\delta}{k}{l}{l}{j}{l}{i}-\Qfracm{\gamma}{\delta}{k}{i}{l}{i}{j}{i}\bigr]\\
+&\lr{\ca{\gamma}c_{\delta}c_{\alpha}\ca{\beta}}_{ijkl}\bigl[\Rfrac{\gamma}{\delta}{k}{j}{k}{i}{l}{k}+\Rfrac{\gamma}{\delta}{i}{j}{i}{k}{l}{i}-\Qfracm{\gamma}{\delta}{l}{j}{l}{i}{l}{k}\bigr]\\
+&\lr{c_{\delta}\ca{\gamma}c_{\alpha}\ca{\beta}}_{ijkl}\bigl[\Rfrac{\gamma}{\delta}{j}{k}{k}{i}{l}{k}+\Rfrac{\gamma}{\delta}{j}{i}{i}{k}{l}{i}+\Qfrac{\gamma}{\delta}{j}{l}{l}{i}{l}{k}\bigr]\\
+&\lr{c_{\delta}c_{\alpha}\ca{\gamma}\ca{\beta}}_{ijkl}\frac{1}{(i\omega-E_{kj})(i\omega-E_{li})}\times\\
&\bigl[\mathcal{R}_{\gamma \delta}(E_{k},E_{l})-\mathcal{R}_{\gamma \delta}(E_{j},E_{i})+\mathcal{Q}_{\gamma \delta}(i\omega, E_{j},E_{l})-\mathcal{Q}_{\gamma \delta}(-i\omega, E_{k},E_{i})\bigr]\\
+&\lr{\ca{\gamma}c_{\alpha}c_{\delta}\ca{\beta}}_{ijkl}\frac{1}{(i\omega-E_{kj})(i\omega-E_{li})}\times\\
&\bigl[\mathcal{R}_{\gamma \delta}(E_{l},E_{k})-\mathcal{R}_{\gamma \delta}(E_{i},E_{j})+\mathcal{Q}_{\gamma \delta}(i\omega, E_{i},E_{k})-\mathcal{Q}_{\gamma \delta}(-i\omega, E_{l},E_{j})\bigr],
\end{split}\label{eq:eq_first_diag_dai}
\end{equation}
with the following definitions for the matrix elements:
\begin{equation}
\lr{\mathcal{O}_{\alpha}\mathcal{O}_{\beta}\mathcal{O}_{\gamma}\mathcal{O}_{\delta}}_{ijkl}=\lr{i|\mathcal{O}_{\alpha}|j}\lr{j|\mathcal{O}_{\beta}|k}\lr{k|\mathcal{O}_{\gamma}|l}\lr{l|\mathcal{O}_{\delta}|i}.
\end{equation}
The functions $\mathcal{R}$ and $\mathcal{Q}$ have the following definitions:
\begin{gather}
\label{eq:R}
\mathcal{R}_{\alpha \beta}(E_{i},E_{j})\Let\frac{1}{Z}(e^{-\beta E_{i}}+e^{-\beta E_{j}})\frac{1}{\beta}\sum_{i\omega'}\frac{\Delta^{\text{d}}_{\alpha \beta}(i\omega')}{i\omega -E_{ij}},\\
\text{\phantom{a}}\nonumber\\
\mathcal{Q}_{\alpha \beta}(i\omega,E_{i},E_{j})\Let\begin{cases}-\frac{\beta}{Z}e^{-\beta E_{i}}\Delta^{\text{d}}_{\alpha \beta}(i\omega)\quad &\text{for }E_{i}=E_{j}\\
\frac{1}{Z}(e^{-\beta E_{i}}-e^{-\beta E_{j}})\frac{1}{\beta}\sum_{i\omega'}\frac{\Delta^{\text{d}}_{\alpha \beta}(i\omega')}{i\omega'- i\omega -E_{ij}}\quad &\text{else.}\end{cases}\label{eq:Q}
\end{gather}
The trace over $\Delta^{\text{D}}$ and $g$ can also be expressed in terms of $\mathcal{R}$:
\begin{equation}
\text{Tr}[g\Delta^{d}]=\sum_{i,j,\delta,\gamma}\lr{i|c_{\delta}|j}\lr{j|c^{\dagger}_{\gamma}|i}\mathcal{R}_{\gamma\delta}(E_{i},E_{j}).
\end{equation}
Hence the analytic continuation of Eq \eqref{eq:sigma_dai} reduces to the continuation of the function $\mathcal{Q}(i\omega,E_i,E_j)$. Since $\mathcal{R}(E_i,E_j)$ does not depend on $i\omega$ the function can be in principle calculated using definition \eqref{eq:R}. In order to treat both function evaluations on the same footing, we discuss in the following how $\mathcal{R}$ and $\mathcal{Q}$ can be calculated via an integral along the real axis.\\
\begin{figure}[t]
\sidecaption
\begin{minipage}[b]{0.64\textwidth}
 \includegraphics[width=\textwidth]{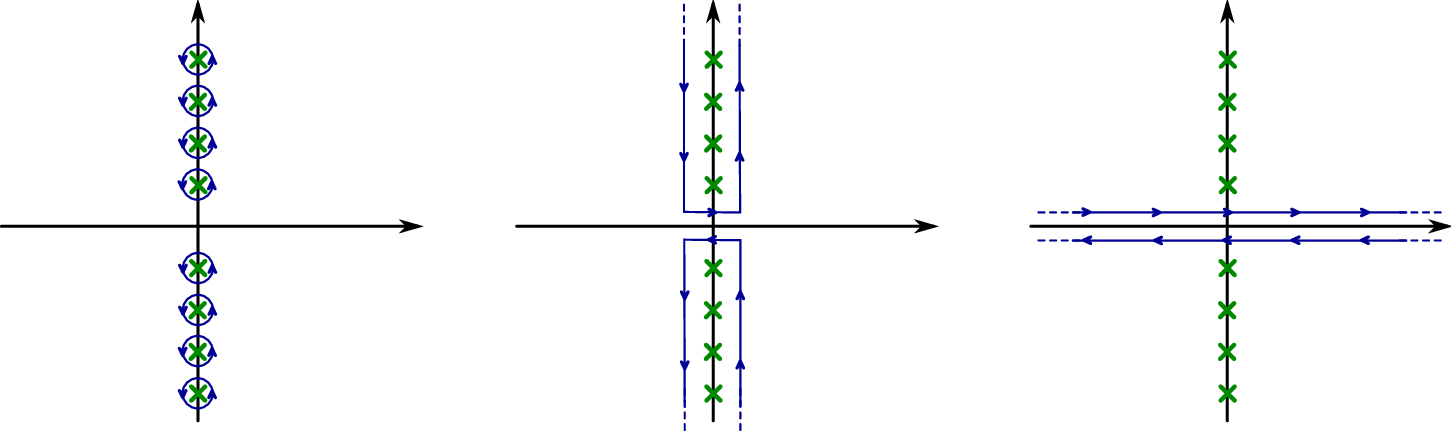}\\[0.5cm]
 \includegraphics[width=\textwidth]{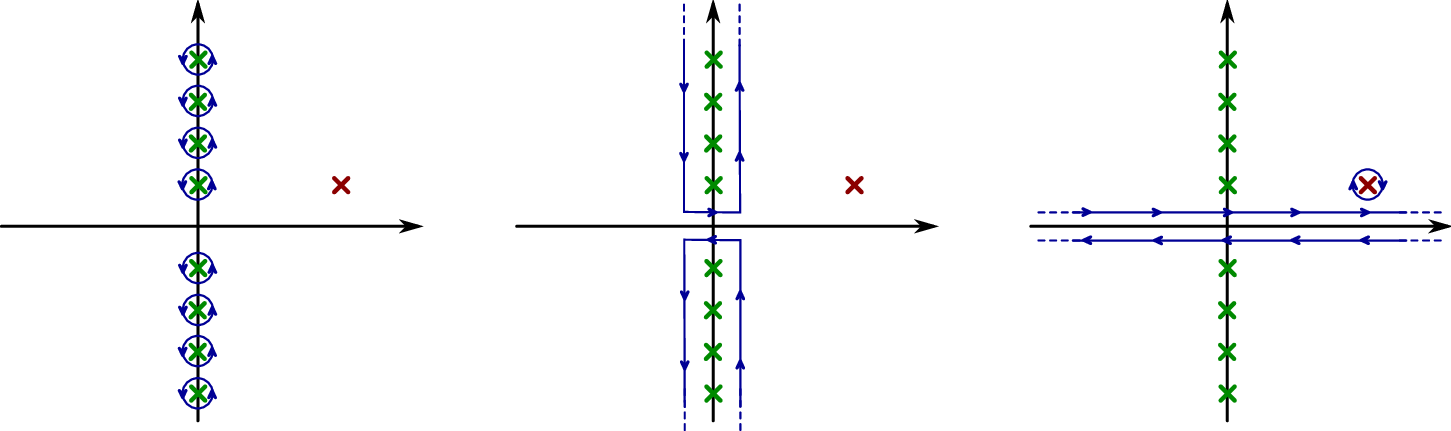}
\end{minipage}
 \caption{\emph{top:} Standard procedure of converting a Matsubara sum into a contour integral. In a first step the sum is replaced by a contour integral over the function itself times the Fermi function, which has poles at Matsubara frequencies. In the second and third step the contour is deformed to two line integrals along the real axis. The illustrated contour deformation is applied in the calculation of $\mathcal{R}$. \emph{bot.:} Construction of the contour integral to calculate $\mathcal{Q}$. In comparison to the upper case one has to take into account an additional pole at $\tilde{z}=i\omega+E_{12}$.}\label{fig:contours}
\end{figure}
The Matsubara sums in $\mathcal{Q}$ and $\mathcal{R}$ are rewritten as a sum over residues of the function itself times the Fermi function, which has poles of first order at the Matsubara frequencies:
\begin{equation}
 \frac{1}{\beta}\sum_{i\omega}F(i\omega)=-\frac{1}{2\pi i}\oint_{C}\frac{F(E)}{e^{\beta E}+1} dE.
\end{equation}
Afterwards the integral contour is deformed according to figure \ref{fig:contours}. This leads to the following definitions of both expressions as integrals along the real axis:
\begin{align}
 \mathcal{R}(E_1,E_2)=&-\frac{X_1+X_2}{2\pi i}\Bigl[\int_{-\infty}^{\infty}\frac{\Delta^{\text{d}}(z^{+})f(z^+)}{z^{+}-E_{12}}-\int_{-\infty}^{\infty}\frac{\Delta^{\text{d}}(z^{-})f(z^-)}{z^{-}-E_{12}}\Bigr]dz,\\
 \begin{split}
 \mathcal{Q}(i\omega,E_1,E_2)=&(X_1-X_2)\Biggl[-\frac{1}{2\pi i}\biggl(\int_{-\infty}^{\infty}\frac{\Delta^{\text{d}}(z^{+})f(z^+)}{z^{+}-i\omega-E_{12}}\, dz\\
&-\int_{-\infty}^{\infty}\frac{\Delta^{\text{d}}(z^{-})f(z^-)}{z^{-}-i\omega-E_{12}}\, dz\biggr)-\frac{1}{1-e^{\beta E_{12}}}\Delta^{\text{d}}(i\omega+E_{12})\Biggr],
\end{split}
\end{align}
where $z^{\pm}=z\pm i\epsilon$, with $\epsilon<\pi/\beta$ being the offset of the contour from the real axis and $X_i=\exp(-\beta E_i)/Z$. The last term in the expression for $\mathcal{Q}$ is due to a residue of $\mathcal{Q}$ at $\tilde{z}=i\omega+E_{12}$. 
Now the analytical continuation can be performed by replacing $i\omega$ by $E+i\delta$:
\begin{multline}
\mathcal{Q}(\tilde{z},E_1,E_2)=(X_1-X_2)\Biggl[-\frac{1}{2\pi i}\biggl(\int_{-\infty}^{\infty}\frac{\Delta^{\text{d}}(z^{+})f(z^+)}{z^{+}-\tilde{z}-E_{12}}\,dz\\
-\int_{-\infty}^{\infty}\frac{\Delta^{\text{d}}(z^{-})f(z^-)}{z^{-}-\tilde{z}-E_{12}}\, dz\biggr)-\frac{1}{1-e^{\beta E_{12}}}\Delta^{\text{d}}(\tilde{z}+E_{12})\Biggr],
\end{multline}
with $\tilde{z}=E+i\delta$, where $\delta$ is the usual broadening parameter, which is restricted to values $\delta>\epsilon$. This completes the analytic continuation.

A few remarks are in place. In contrast to Ref. \cite{Dai:2005qd} we do not perform the limit $\epsilon\to 0$ because of the particular structure of the ``dual'' hybridization, which has poles on the real axis. The integrals are evaluated with a small offset $\epsilon$. For not too low temperatures a simple quadrature rule is sufficient.
For lower temperatures, $\Delta^{\text{d}}$ develops a sharp peak. We therefore employ an adaptive Gauss-Kronrod algorithm taken from the GNU scientific library \cite{1538674}. As a check for numerical accuracy one may verify that the value of the integrals is independent of $\epsilon$.\\

\begin{figure}[t]
\sidecaption
\begin{minipage}[b]{0.6\linewidth}
\begin{minipage}{0.49\linewidth}
\centering
\includegraphics[width=\textwidth]{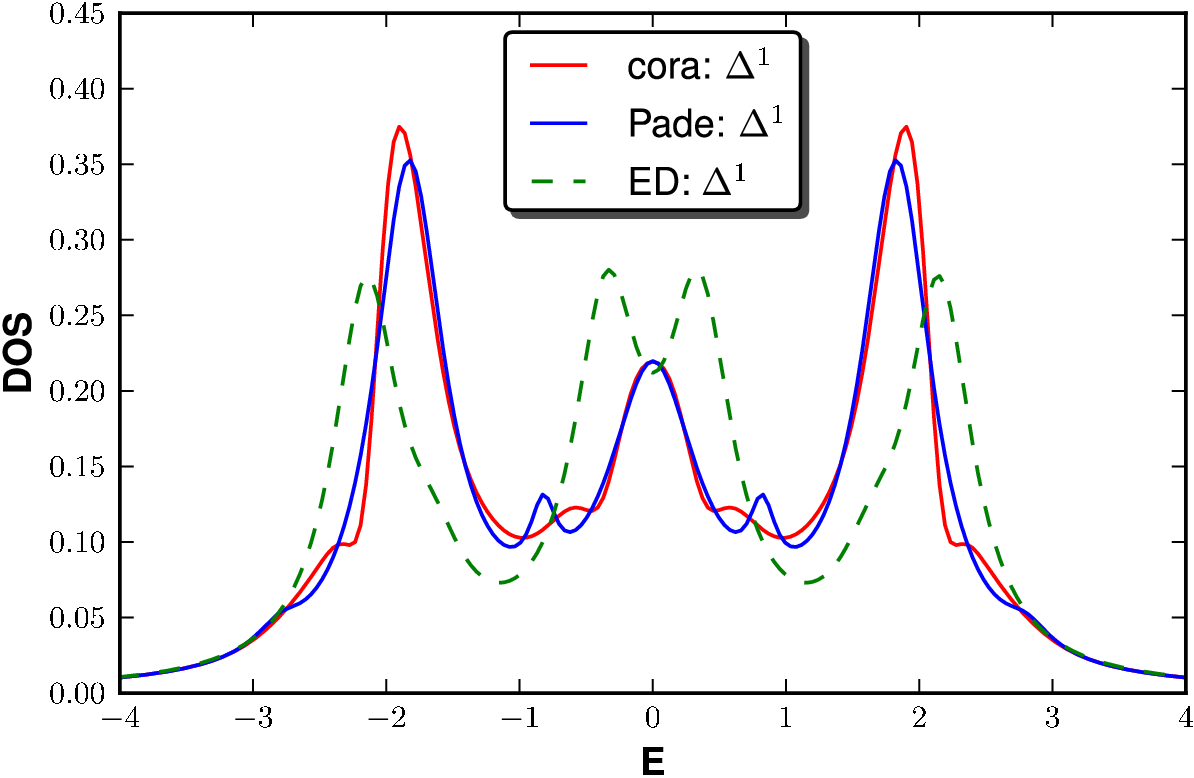}
\end{minipage}
\hspace{0.01cm}
\begin{minipage}{0.49\linewidth}
\centering
\includegraphics[width=\textwidth]{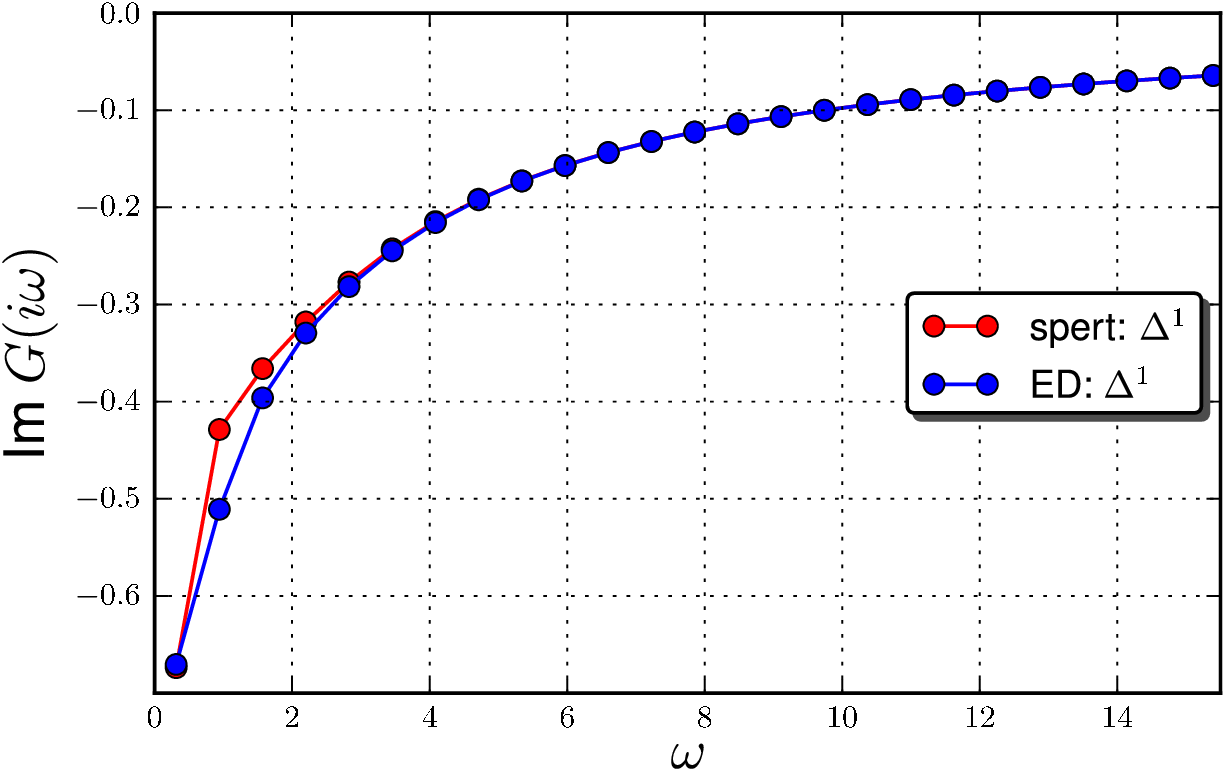}
\end{minipage}
\\[0.3cm]
\centering
\begin{minipage}{0.49\linewidth}
\centering
\includegraphics[width=\textwidth]{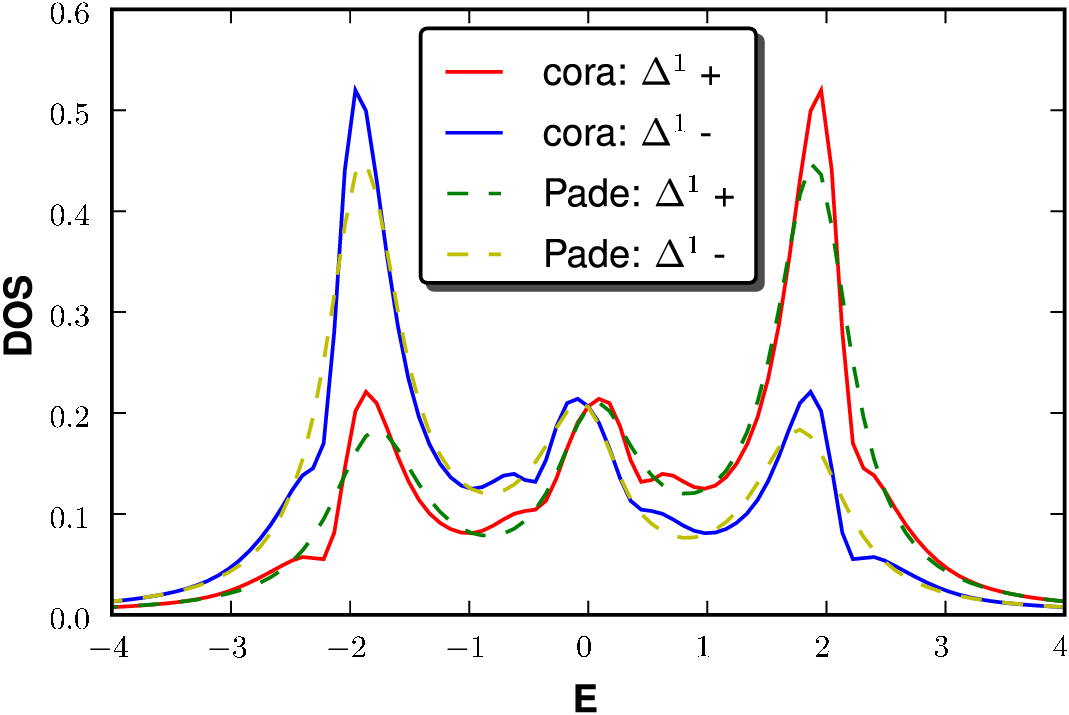}
\end{minipage}
\hspace{0.01cm}
\begin{minipage}{0.49\linewidth}
\centering
\includegraphics[width=\textwidth]{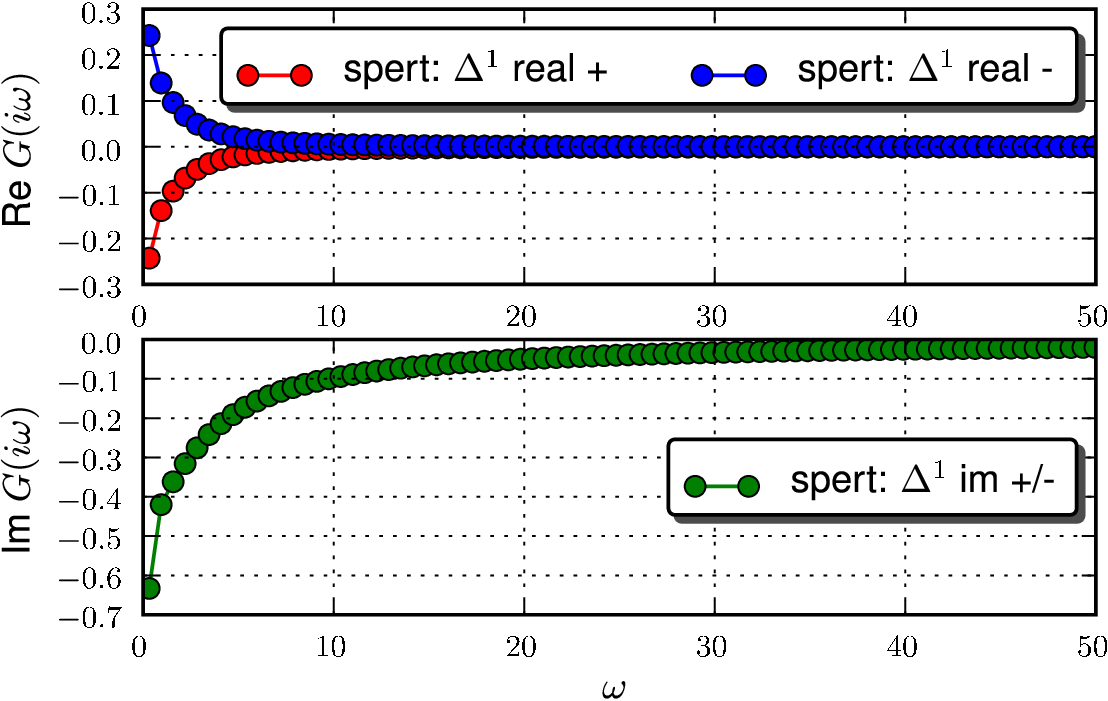}
\end{minipage}
\end{minipage}
\caption{SPERT calculations on the real axis (CORA), in comparison with Pad\'e and the result of the reference system (ED). The upper part shows an example for $\beta=10$ and $U=3$, in units of the half-bandwidth $(W/2)$. Here the agreement between CORA and Pad\'e is good, but the CORA shows more structure on the real axis. The lower part shows the same system with an applied magnet field: $B=0.05$.
}
\label{fig:spert_real_example}
\end{figure}
Figure \ref{fig:spert_real_example} shows some illustrative results for the AIM with hybridization corresponding to a semielliptical DOS of bandwidth $W$. In the upper left plot the calculation on the real axis (CORA) is compared to an analytic continuation using Pad\'e and the initial solution of the reference system (labeled 'ED'). The ED curve has a clear splitting at the Fermi level, whereas the CORA curve exhibits a Kondo peak. The CORA is in a good agreement with Pad\'e. The lower left plot shows an additional example with an applied magnetic field. Here the splitting of the peaks is clearly visible and the CORA is again in good agreement with Pad\'e. For completeness the data on Matsubara frequencies has been added on the right.\\
\begin{figure}[t]
\sidecaption
\begin{minipage}[b]{0.6\textwidth}
 \includegraphics[width=\textwidth]{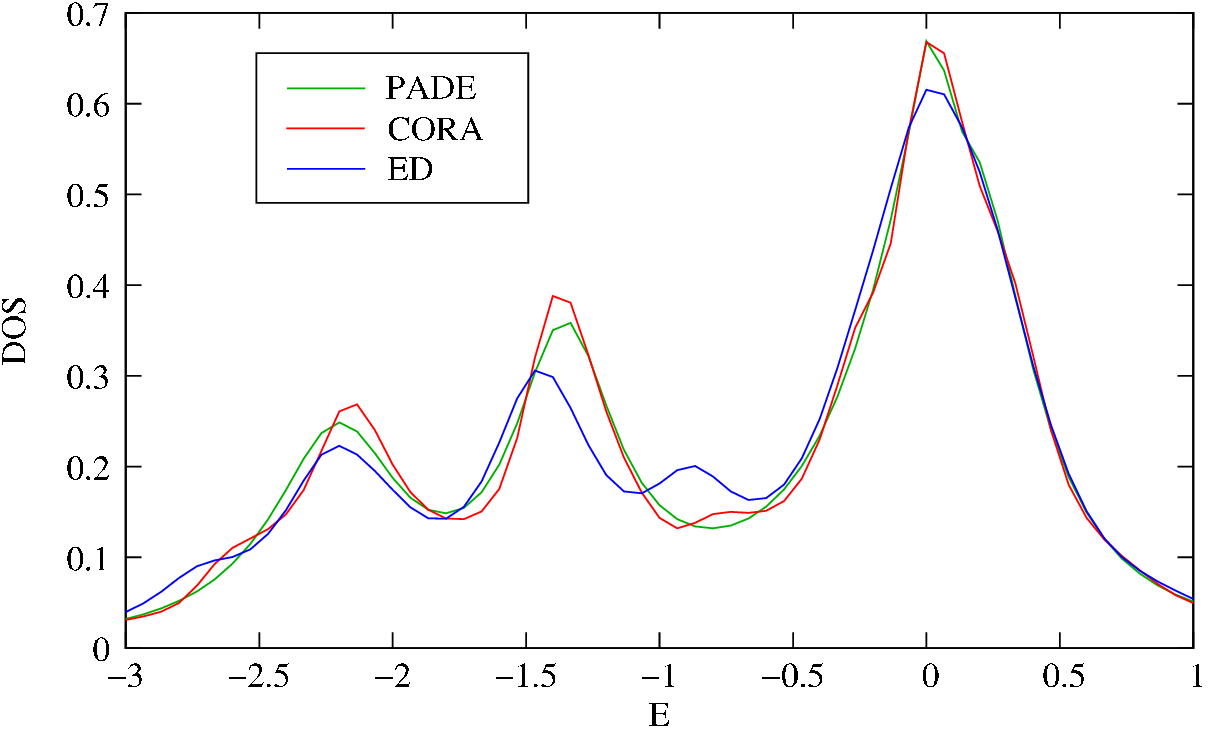}
\end{minipage}
 \caption{Example of a CORA calculation for a multi-orbital case. A three orbital impurity with $U=1.5$, $U'=0.7$, $J=0.4$, $\mu=3.02$ and $\beta=15$ (in units of the quarter-bandwidth) is embedded in a bath with a constant density of states in the energy window $|E|<W/2$. The coupling to the bath is $V=0.1$. The hybridization was approximated by a single bath site. The CORA data set exhibits small features in the density of states which are absent in the Pad\'e results.}\label{fig:multiorbital}
\end{figure}
In Figure \ref{fig:multiorbital} we present an example for a multi-orbital problem with a rotational invariant Coulomb vertex. A three-orbital impurity has been embedded in a bath, which corresponds to a flat density of states in the energy window $|E|<W/2$. The coupling to the bath was moderate and has been approximated by a single bath site, which was equally connected to all impurity orbitals.
In comparison to the solution of the reference system a clear shift of nearly all peaks in the CORA is visible. The Pad\'e solution is in good overall agreement with the CORA data, but fails to reproduce some small features. 
Indeed one may expect that the CORA results are more accurate than the analytical continuation via Pad\'e approximants if the continued function has rich structure, since the approximation through a rational function becomes less accurate.
For complicated multiorbital systems we expect differences to be more pronounced.

\section{Conclusion}
In the present work we have presented a multiorbital impurity solver based on the superperturbation for the Anderson impurity model (AIM). It allows to compute dynamical quantities directly on the real axis, which has the advantage that no analytic continuation using approximate methods like MAXENT or Pad\'e is necessary. Considering three examples including a multi-orbital impurity problem, we compared the results of the new implementation to results obtained using Pad\'e. 
We find overall good agreement. The direct calculation on the real axis however can resolve finer structures, which will be useful for the study of multiplett effects in solids.
This work has been supported by the Cluster of Excellence Nanospintronics (LExI Hamburg) and DFG Grant(436113/938/0-R).
\bibliographystyle{adp}
\providecommand{\WileyBibTextsc}{}
\let\textsc\WileyBibTextsc
\providecommand{\othercit}{}
\providecommand{\jr}[1]{#1}
\providecommand{\etal}{~et~al.}

\end{document}